\documentclass[12pt]{iopart}
\usepackage[english]{babel}
\usepackage{
amssymb,amsfonts,latexsym}
\usepackage[pdftex]{graphicx}
\usepackage{color}

\usepackage{times}
\usepackage[colorlinks,bookmarks=false,citecolor=blue,linkcolor=red,urlcolor=blue]{hyperref}

\def\HH{\mathcal H}

\begin{document}

\title{Entanglement evolution across defects in critical anisotropic Heisenberg chains}

\author{Mario Collura and Pasquale Calabrese}
\address{Dipartimento di Fisica dell'Universit\`a di Pisa and INFN, 56127 Pisa, Italy}
\date{\today}

\begin{abstract}
We study the out-of-equilibrium time evolution after a local quench connecting two anisotropic spin-1/2 XXZ Heisenberg open chains 
via an impurity bond. The dynamics is obtained by means of the adaptive time-dependent density-matrix renormalization group. 
We  show that  the entanglement entropies (Von Neumann and R\'enyi), 
in the presence of a weakened bond depend on the sign of the bulk interaction. 
For attractive interaction ($\Delta<0$), the defect turns out to be irrelevant and the evolution is asymptotically 
equivalent to the one without defect obtained by conformal field theory. 
For repulsive interaction ($\Delta>0$), the defect is relevant and the entanglement saturates to a finite value. 
This out-of-equilibrium behavior generalizes the well known results for the ground-state entanglement entropy of the model. 
\end{abstract}

\section{Introduction}

Recent times witnessed a renewed interest in the non-equilibrium dynamics of isolated many-body quantum systems.
A particular class of these non-equilibrium problems which is experiencing a dramatic explosion of theoretical 
activity is that of a  sudden quench of a Hamiltonian parameter, principally boosted  by 
the experiments on trapped ultra-cold atomic gases
\cite{uc,kww-06,tc-07,tetal-11,cetal-12,getal-11,shr-12,rsb-13} in which it has been shown that
it is possible to follow the  unitary non-equilibrium 
evolution without any significant coupling to the environment. 
These experiments fall into two main classes which are usually 
denoted as global and local quantum quenches. 
In the former case, a control parameter is suddenly quenched in the whole 
system (usually in a translationally invariant manner), while in the latter 
the change is only local, see e.g. \cite{revq} for a review.
Global quantum quenches are ideal experiments to investigate the intriguing issue 
of the existence of a stationary state and thermalization \cite{cc-06,gg,cdeo-08,bs-08,CEF,eef-12,se-12,bdkm-11,fm-10}, while local quenches 
usually reveal the spreading of information and correlations in a cleaner way because they mainly probe 
universal low-lying excitations.   
Other interesting effects can be uncovered in some `intermediate' non-equilibrium situations, 
such as inhomogeneous quenches \cite{chl-08,sc-08,eip-09,dra,mpc-10} and 
gas expansion \cite{shr-12,rsb-13,mg-05,v-12,ck-12,hm-v}. 

In this paper we consider the most commonly studied local quench in which two halves of a {\it critical} one-dimensional 
system are initially prepared in their respective ground states and at a given time,
let us say $t=0$, they are connected and let evolve according to a unitary Hamiltonian dynamics. 
Previous studies \cite{cc-05,cc-07l,lk-08,fc-08,ep-08,edpp-08,isl-09,ds-11a,ds-11,isl-12,lbb-12} 
showed that a very useful tool to understand the  non-equilibrium quench dynamics is 
the entanglement entropy between two complementary parts.  
For a general bipartition of a pure state $|\Psi\rangle$ of quantum system (i.e. writing the whole Hilbert space of the system 
 as a direct product of two parts $\HH=\HH_A\otimes\HH_B$),
the R\'enyi entropy of the reduced density matrix $\rho_{A}={\rm Tr}_B |\Psi\rangle\langle\Psi|$ 
of the subsystem $A$ \cite{rev}
\begin{equation}
S^{(\alpha)}_{A} = \frac{1}{1-\alpha}\ln\mathrm{Tr}\rho^{\alpha}_{A},
\end{equation}
is a proper measurement of the entanglement between the two parts. 
In the limit $\alpha\to 1$, $S^{(1)}_{A}$ reduces the more studied von Neumann entanglement entropy
but, the knowledge of the R\'enyi entropies for any $\alpha$ gives far more 
information than the $\alpha=1$ case because it provides  
the full spectrum of the reduced density matrix \cite{cl-08}.

In the ground-state of a one-dimensional critical system, whose scaling limit is described by a conformal
field theory (CFT), in the case when $A$ is an interval of length $\ell$ 
embedded in an infinite system, the asymptotic large $\ell$ behavior of the R\'enyi
entropies is  given by \cite{Holzhey,cc-04,Vidal,cc-rev}
%
\begin{equation}
S^{(\alpha)} = \frac{c}{6}\left( 1 + \frac{1}{\alpha}\right)\ln\ell + c'_{\alpha},
\end{equation}
where $c$ is the central charge  \cite{c-lec} and $c'_{\alpha}$ a non-universal constant. 
In the case when the whole system is finite, the R\'enyi entropies can be obtained by a conformal 
mapping from the cylinder to the plane and the net effect is just to replace $\ell$ with the chordal length 
$L/\pi \sin(\pi\ell/L)$ \cite{cc-04}.  
More complicated and model dependent universal expressions for $S^{(\alpha)}$
have been  obtained for the low-lying excited states conformal theories \cite{abs-11,elc-13} as well.

Conformal invariance can also be exploited to predict the behavior of the entanglement entropy for a 
`cut and glue' local quench when two semi-infinite critical systems are joined  
together in a translational-invariant/homogeneous way (i.e. in a microscopical model, 
the added coupling is equal to all the others in the system).
It has been found \cite{cc-07l} that the entanglement grows logarithmically with time, 
again with a prefactor given by the central charge, i.e. 
\begin{equation}
S^{(\alpha)} = \frac{c}{6}\left( 1 + \frac{1}{\alpha}\right)\ln t + {\rm cst.}
\end{equation}
This has been generalized by St\'ephan and Dubail \cite{ds-11} to the case when the two parts 
which are connected at $t=0$ have {\it finite} length. 
In the special case when the two parts have the same length $L/2$, the final result is 
\begin{equation}
S_{CFT}^{(\alpha)} = \frac{c}{6}\left( 1 + \frac{1}{\alpha}\right)\ln\left|\frac{L}{\pi}\sin\left(\frac{\pi v t}{L}\right) \right| + {\rm cst.},
\label{ds-form}
\end{equation}
where we explicitly introduced the speed of the sound $v$.
These results, as well as some for general observables, have been largely tested 
in the  literature both for free and interacting models \cite{ep-08,edpp-08,isl-09,ds-11a,ds-11,gkss-05,s-08,dir-11,gree-12,ss-12}. 
Furthermore it is worth mentioning that a few proposals to measure the entanglement entropy in real experiments
are based on local quantum quenches \cite{kl-08,hgf-09,c-11,ad-12}. 

It is natural to wonder how the previous equilibrium and non-equilibrium results generalize to the case 
when the two halves are joined with a coupling that is not equal to the others, especially in view
of their experimental realizations, where the control of the added bond can be not perfect. 
The simpler case of non-interacting fermions turned out to be very peculiar and not 
general enough, as opposite to the homogenous case. 
Indeed, it has been found that the defect is marginal \cite{kane1992,eggert1992}. 
For the ground-state entanglement entropy this implies that the entanglement between the two parts separated by the 
defect still grows logarithmically with the system size $L$, but with a pre-factor $C_{\alpha }$ that 
continuously interpolates as function of the defect strength between $c/6(1+1/\alpha)$ in the absence of the defect and $0$ when
the defect is so strong to divide the system in two parts. 
This effect was firstly found numerically \cite{peschel2005,isl-09} and only later the interpolating 
function has been exactly calculated as function of the defect strength \cite{ep-10,cmv-12,peschel2012}, which for 
general $\alpha$ reads
\begin{equation}\fl
C_{\alpha }(s) = \frac{2}{\pi^2 (1-\alpha)} 
\int_0^\infty d x \ln \left [\frac{1+\e^{-2 \alpha \omega(x,s)}}
{\left(1+\e^{-2\omega(x,s)}\right )^\alpha }\right ]  , \qquad
\omega (x,s) = {\rm acosh} \left [\frac {\cosh (x)}{s}\right ] ,  
\label{e9}
\end{equation}  
where $s\in[0,1]$ is the transmission amplitude simply related to the defect strength (see below).
Finally, very recently, it has been shown that the same interpolating function enters also in the prefactor 
in Eq. (\ref{ds-form}) for the time evolution of the entanglement entropy after a local quench connecting two 
chains of free spinless fermions \cite{eisler2012,bcm-prep}.
All these results for non-interacting fermions in the presence of a defect still await a full CFT derivation which, also in view of the 
similar treatment in Ref. \cite{ss-08}, must be clearly possible.


For bulk {\it interacting} spinless fermions, the asymptotic behavior of the entanglement entropy is completely different.   
Indeed, in {\it equilibrium} an old renormalization group (RG) analysis for one-dimensional interacting fermions 
(modeled as a Luttinger liquid field theory) shows that the relevance of the defect depends on the sign of the 
bulk interaction \cite{kane1992,eggert1992}. 
For attractive interaction, the defect turned out to be irrelevant, i.e. for large enough distances, 
the system behaves like if the defect was not present, while for repulsive interaction the defect is 
relevant and even a small defect cuts the chain into two almost independent halves. 
Also the logarithmic behavior of the entanglement entropy was calculated in Ref.  \cite{levine2004} to the lowest order in 
the impurity strength. 
Finally, a numerical study of the spin-$1/2$ anisotropic Heisenberg XXZ chain 
with a central defect was presented in Ref. \cite{zhao2006} and 
it has been found that the von Neumann entanglement entropy grows logarithmically with $L$ 
for $\Delta < 0$ (ferromagnetic region corresponding to attractive interaction) with a prefactor equal to $1/3$
(i.e. confirm the irrelevance of the defect because $c=1$) 
and instead saturates for $\Delta > 0$ (antiferromagnetic region corresponding to repulsive interaction), 
confirming the relevance of the impurity. 
The approach to the thermodynamic limit is non-universal and obviously depends on the defect strength.
The same qualitative results have been also found for a spin-1 gapless model \cite{rzh-09}.


\begin{figure}[t]
\center\includegraphics[width=\textwidth]{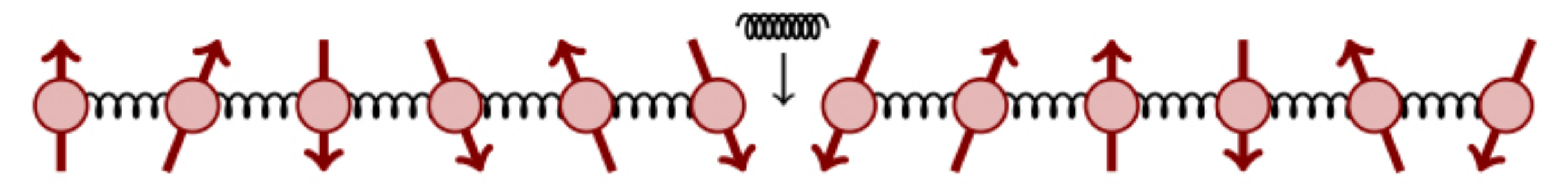}
\caption{\label{fig0} 
Graphical representation of the local quench considered in this paper: 
a spin-chain is prepared in the ground-state of two disconnected halves
that at time $t=0$ are joined through a defect bond (represented as a spring different from the others).} 
\end{figure}

In this paper we 
study the time evolution of  the anisotropic XXZ Heisenberg chain 
after a local quench connecting two initially separated half-chains with a 
defect bond, as shown in Fig.~\ref{fig0}. 
We tackle the problem considering the evolution of an open chain with one modified bond in the center 
and using the adaptive time-dependent density matrix renormalization group (tDMRG) \cite{tDMRG}. 
This numerical method is well suited for the calculation of the entanglement entropy 
since at each step of the algorithm one needs to reconstruct 
the reduced density matrix and its spectrum. 
The manuscript is organized as follows.
In Sec. \ref{Sec2} we present the details of the model and of the numerical method. 
In Sec. \ref{Sec3} we report the time evolution of the entanglement (von Neuman and R\'enyi) entropy. 
The simulations have been performed for different values of the defect strength $\delta$ and anisotropic parameter $\Delta$,
as well as for a few chain lengths $L$. 
Finally in the last section we draw our conclusions.

\section{The model, the local quench, and the method}
\label{Sec2}

We consider the anisotropic XXZ Heisenberg chain with $L$ sites and a defect $\delta$ in the center 
defined by the Hamiltonian 
\begin{equation}
H = \sum_{i=1}^{L-1} J_{i}\left[S^{x}_{i}S^{x}_{i+1} +  S^{y}_{i}S^{y}_{i+1} + \Delta S^{z}_{i}S^{z}_{i+1}\right],
\label{HXXZ}
\end{equation}
with 
\begin{equation}
J_{i} = \left\{\begin{array}{ll}1, & i\neq L/2, \\1+\delta, \qquad& i=L/2.\end{array}\right.
\label{Jiev}
\end{equation}
and we impose open boundary conditions on the sites $i=1$ and $i=L$. 
Here $S^{\alpha}_{i}$ are the local spin-$1/2$ operators, i.e. in terms of Pauli matrices
$S^{\alpha}_{i}=\sigma^{\alpha}_{i}/2$.
In the homogenous case with $\delta=0$, 
the model is integrable by means of Bethe ansatz \cite{KorepinBOOK} allowing 
for an exact characterization of the ground state and all excited states. 
For $|\Delta|\leq 1$ the model is gapless, conformal (for $\Delta\neq-1$) and described by a Luttinger 
liquid field theory with central charge $c=1$ and Luttinger parameter $K=\pi/(2\arccos \Delta)$, while for $|\Delta|>1$ it acquires a gap.  
For $\Delta\neq0$, although the model is Bethe ansatz solvable, it is still not known how to use integrability
to calculate effectively the entanglement entropies in the gapless phase, in spite of   several attempts in the 
literature \cite{XXZ-var} (in the gapped phase instead some exact results are known \cite{cc-04,gap}). 

Via a Jordan-Wigner transformation, apart from boundary terms,
the chain in Eq. (\ref{HXXZ}) is mapped into a lattice model of spinless fermions $c_j$ (which satisfy canonical anti-commutation 
relations $\{c_l,c^\dagger_m\}=\delta_{l,m}$)
with Hamilltonian
\begin{equation}
H= \sum_{i=1}^{L-1} \frac{J_{i}}2\left[c^\dagger_{i}c_{i+1} +  c^\dagger_{i+1}c_{i} + 
2{\Delta} \Big(c^\dagger_{i}c_{i}-\frac12\Big)\Big(c^\dagger_{i+1}c_{i+1}-\frac12\Big)    \right],
\end{equation}
which has attractive nearest-neighbor interaction for $\Delta<0$ and repulsive for $\Delta>0$.
The case $\Delta=0$ corresponds to free spinless fermions (or XX chain). 

In this manuscript we consider the non-equilibrium situation in which 
the chain is initially prepared in the relative ground states of two equal parts of length $L/2$, i.e. 
in the global ground-state of the Hamiltonian (\ref{HXXZ},\ref{Jiev}) with $\delta_0=-1$.
We fix $L$ to be an integer multiple of $4$, so that the ground state of each of the two initially separated 
parts is non-degenerate and has zero magnetization (i.e. the corresponding fermion lattice is half-filled).
At time $t=0$, these two parts are connected by a defect bond $J_{L/2}=J_{\rm def}=1+\delta$ characterized by a deficit $\delta$, 
while the rest of the chain is left unchanged with $J=1$. 
The transmission amplitude $s$ in Eq. (\ref{e9}) is related to the defect strength as $s=\sin(2\arctan (1+\delta))$ \cite{ep-10}, 
for $-1\leq\delta\leq0$. 
The above assumptions are not only technical: away from half-filling a marginal operator 
(corresponding to forward scattering in fermion language) can be present and it is expected to modify 
the behavior of the entanglement entropies as it is known in equilibrium for other observables \cite{al-94,az-97}. 
However, a proper analysis of these cases is beyond the goal of this paper.

Using the general CFT result reported in the introduction, for a homogeneous quench (i.e. with $\delta=0$)
the von Neumann entanglement entropy (hereafter we will refer to $S^{(1)}$ just as $S$ to enlighten the notation)
is \cite{ds-11}
\begin{equation}\label{S_CFT}
S_{CFT} = \frac{1}{3}\ln\left|\frac{L}{\pi}\sin\left(\frac{\pi v  t}{L}\right) \right|,
\end{equation}
and also the anisotropy strength dependence of the sound velocity is exactly known \cite{KorepinBOOK}
\begin{equation}\label{v_F}
v=\frac{\pi\sin(\arccos\Delta)}{2\arccos\Delta},
\end{equation}
which is valid in the gapless phase $\Delta\in[-1,1]$.
Notice that Eq. (\ref{S_CFT}) is periodic in time with period $t_r=L/v$. We will refer to $t_r$ as {\it revival time}.
The existence of the revival time is a manifestation of the well-known light-cone effect \cite{cc-05,cc-06,cc-07l,ds-11}.
Indeed $t_r$ is  the time needed for an excitation moving with speed $v$ 
starting from the center to arrive to the boundary, being reflected elastically and return to the center. 
In a local quench and in the thermodynamic limit, since the excess of energy of the initial state 
(compared to the ground state) is intensive, only low-lying universal excitations can be populated
and these have all the same speed $v$.
Clearly, in systems with a finite number of degrees of freedom, quasi-particles with different velocities
could be excited even by a small energy excess resulting in imperfect revivals
(the precise interference due to finite size-scaling effects in conformal systems is largely discussed  in Ref. \cite{ds-11}
to which we remand for concrete examples). 
Notice that since $v$ in Eq. (\ref{v_F}) goes to zero approaching $\Delta=-1$, then $t_r$
becomes very large and no revival will be numerically accessible in this limit.
However, for $\Delta=-1$ the model is not conformal invariant anymore because the 
dynamical critical exponent becomes $z=2$.


\begin{figure}[t]
\center\includegraphics[width=0.7\textwidth]{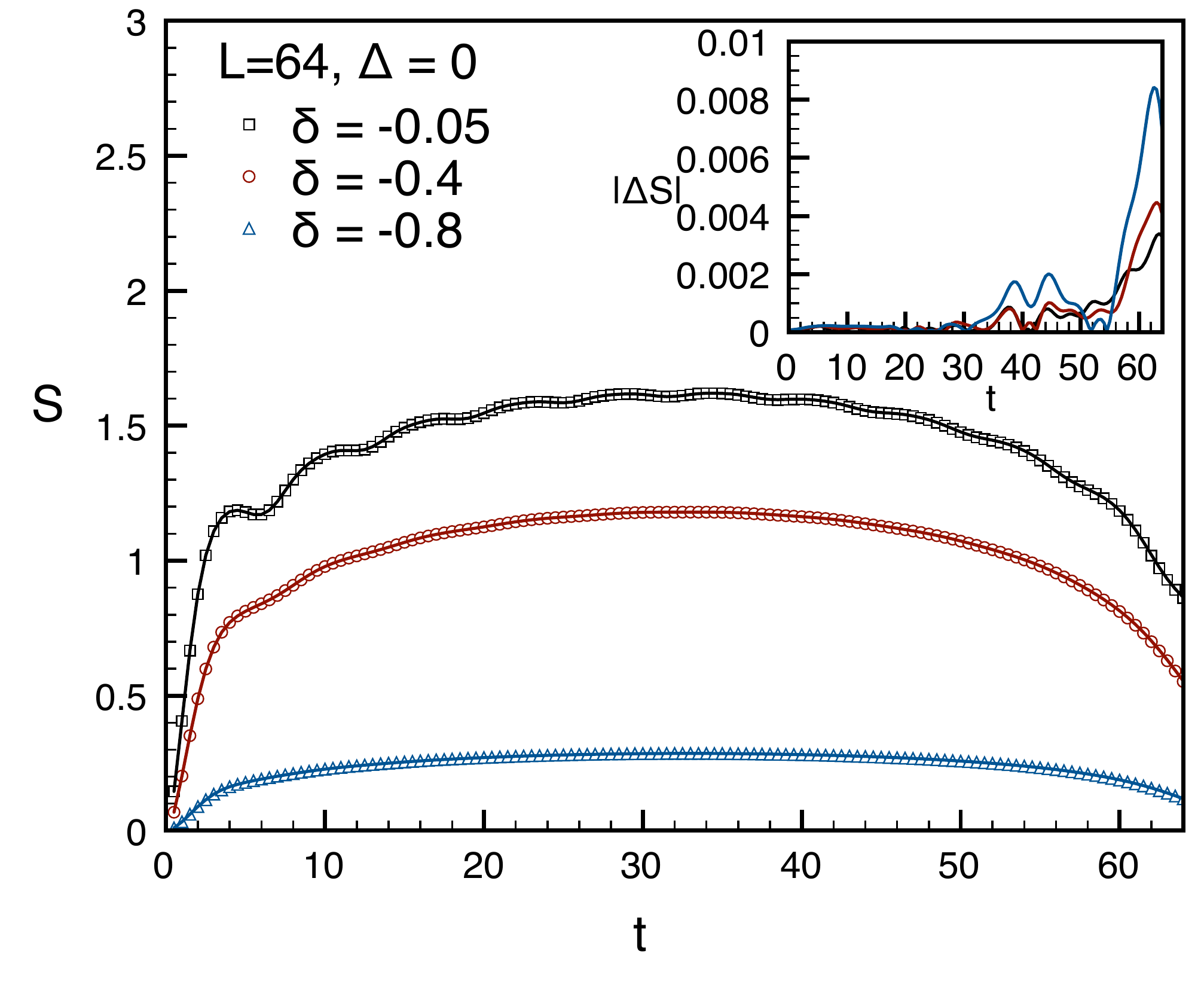}
\caption{\label{fig1} 
Comparison of the entanglement entropy between tDMRG (symbols) and exact results (lines) in the 
XX chain ($\Delta=0$) for three different impurity strengths. 
The data in the main panel agree perfectly. 
The inset shows the relative difference $|\Delta S| \equiv |S_{\rm tDMRG}-S_{\rm ex}|/S_{\rm ex}$. 
The relative error is smaller than $0.2\%$ for almost the entire duration of the temporal evolution. 
The final increase is mainly due to the decreasing of the absolute value of the entropy which therefore 
implies an increase of the relative error. Nevertheless, the relative error is always less than $1\%$.  } 
\end{figure}

\subsection{Method}

We study the local quench dynamics of the XXZ chain by means of extensive tDMRG simulations \cite{tDMRG}.
The algorithm initially performs a static subroutine \cite{DMRG} which selects the initial state as the tensor product of the 
non-degenerate ground states of the two halves. In the decimation process, 
we keep a number of states such that the energy precision is at least of the order of $10^{-12}$. 
Subsequently, we perform the evolution in the presence of the impurity bond using the time-adapting block-decimation 
procedure implemented in tDMRG code \cite{tDMRG}. 
We use the second order Suzuki-Trotter decomposition of the evolution operator with time step $dt=5\cdot10^{-2}$. 
For each time step, the local evolution operator is applied sequentially on each bond starting form the left boundary of the chain. 
We adapt in time the number of states used to describe the reduced Hilbert space retaining at each local step 
all those eigenvectors of the reduced density matrix corresponding to eigenvalues larger than $10^{-16}$, 
up to a maximum value $\chi_{\rm MAX}=200$ 
(clearly the effective maximal value used by the algorithm strongly depends on the simulation parameters). 
As well known (se e.g. Refs. \cite{Vidal,tns}), the computational complexity of the time evolution of a quantum system on a 
classical computer using the tDMRG algorithm is essentially set by the growth of the bipartite entanglement. 
As the entanglement increases with time, we have to enlarge the dimension $\chi$ of the reduced 
Hilbert space in order to optimally control the truncation error. 
In spite of this refined adaptive choice of $\chi$, the truncation procedure is the main source of error of the algorithm for the 
largest system sizes, but if we would let $\chi$ to grow without restrictions, the algorithm would easily get stuck 
in never ending computations.
In the following we will only consider the case of a weakened bonds, 
i.e. $\delta\in[-1,0]$ in Eq. (\ref{Jiev}). 

To test the accuracy of the algorithm, we benchmarked it in the XX case ($\Delta=0$) 
which is a simple free-fermion hopping model and wherein one can exactly calculate the finite-size entropy from the 
one-particle correlation matrix $\langle c^{\dag}_{i}c_{j} \rangle$ \cite{peschel2003,Vidal}, 
even for a chain with defects \cite{peschel2005}. 
In Fig. \ref{fig1} we report the numerical results for the time evolution of the entanglement entropy in a XX chain 
of $L=64$ spins after a local quench with a defect bond.
In the figure, we confront the tDMRG data with the exact numerical calculations for three different values of the 
defect strength $\delta$. 
The relative error  increases with time as expected, but it remains 
bounded below $1\%$ (in this case, 
this error is essentially due to the accumulation of the truncation errors).

\section{Time evolution of the entanglement entropy}
\label{Sec3}

By means of tDMRG, we calculate numerically the time evolution of the entanglement entropies (von Neumann and R\'enyi)
for spin-chains of length $L=32, 64,128$. 
For the smaller system sizes $L=32,64$ we consider many possible values of the defect strength, namely 
all values of $\delta=-0.1 m$ with $m=0,1,\dots, 9$ and $\delta=-0.05$,  
and several values of  the bulk anisotropy parameter $\Delta\in(-1,1]$, namely all $\Delta=0.1 m$ with $m=-9,-8\dots 10$
(i.e. only values of $\Delta$ such that the scaling limit of the chain is described by a CFT with $c=1$).

\begin{figure}[t]
\includegraphics[width=0.5\textwidth]{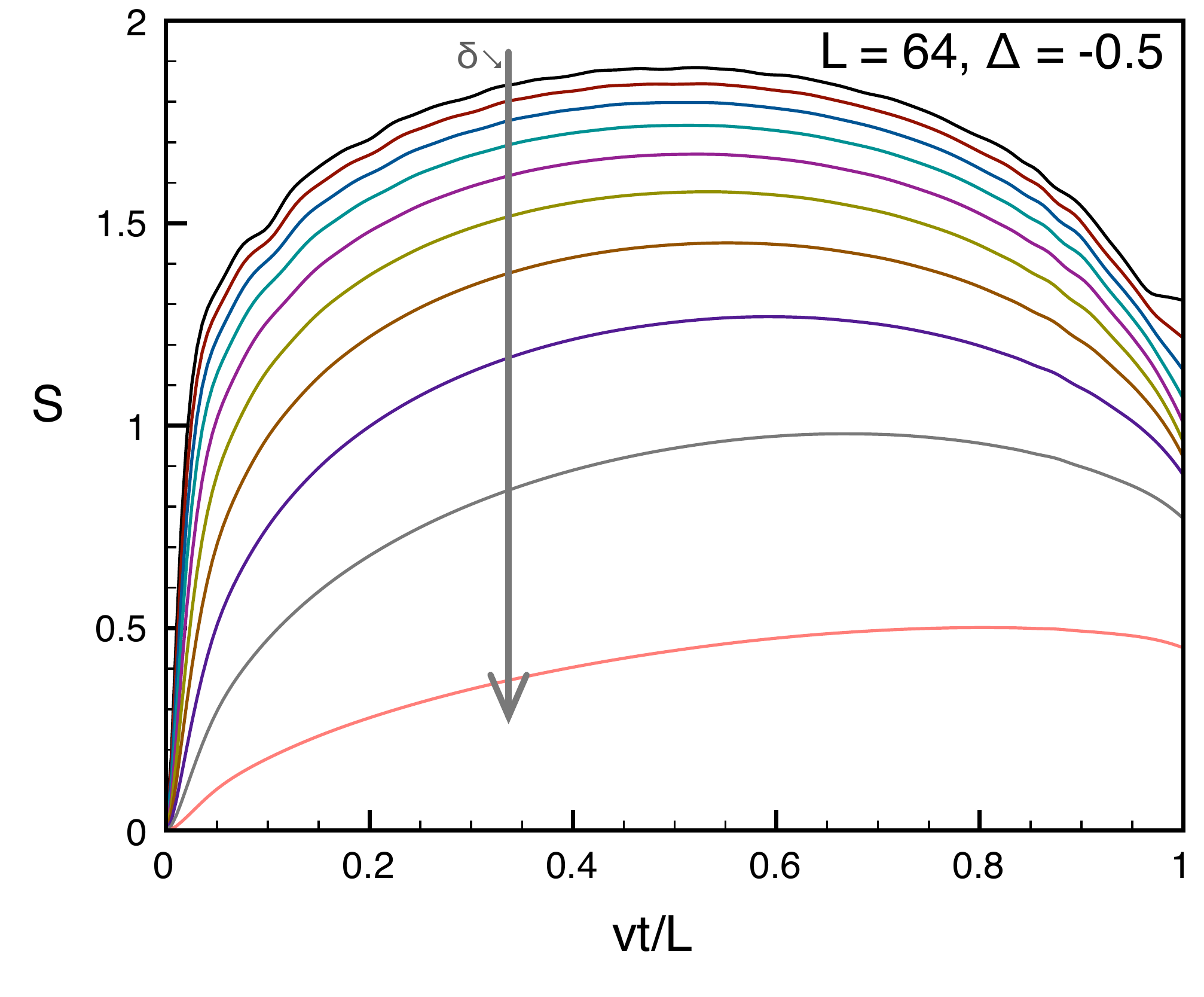}\includegraphics[width=0.5\textwidth]{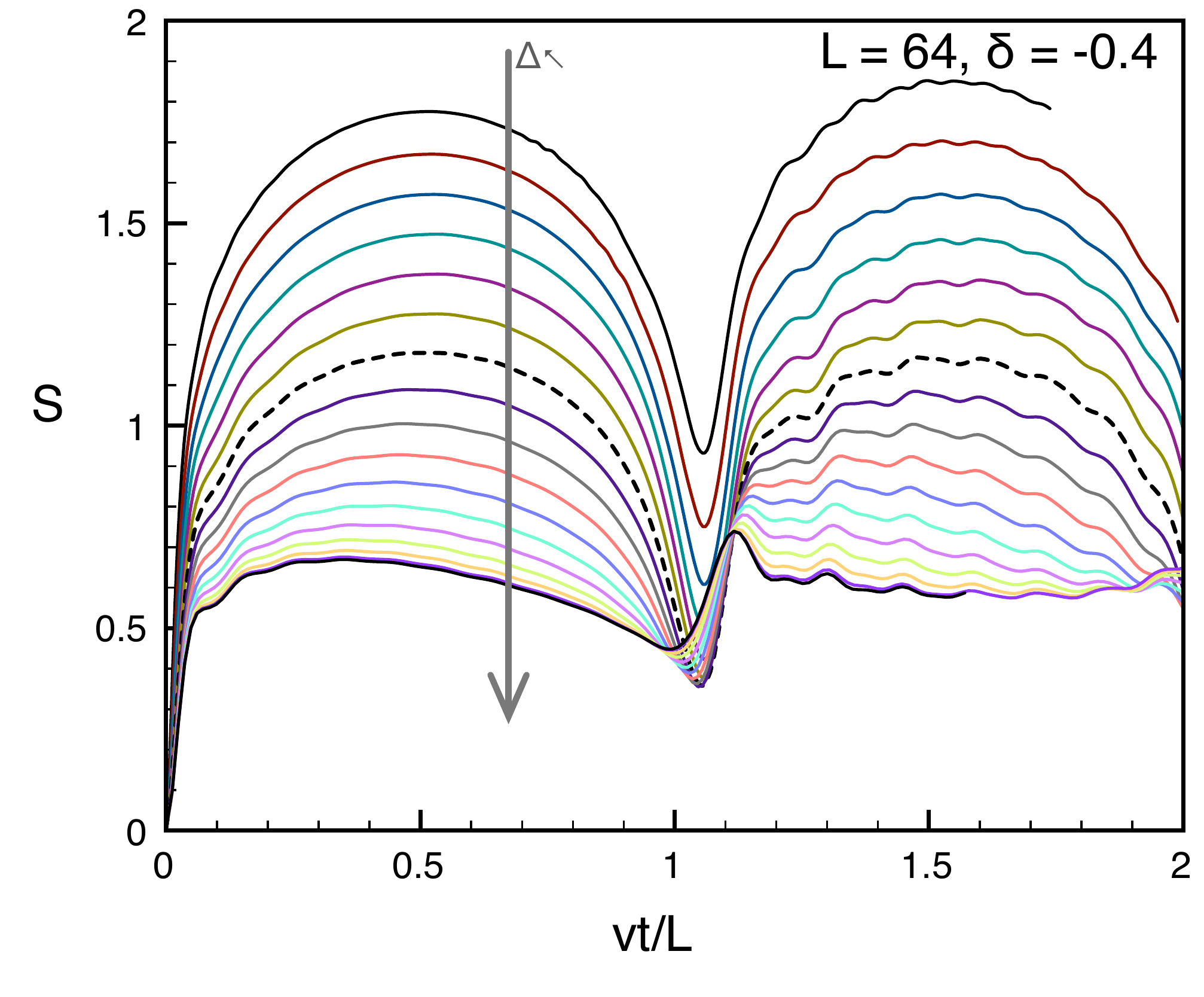}
\caption{\label{fig64} 
Time evolution of the entanglement entropy after connecting two XXZ spin chains with total length $L=64$. 
Left: We report the results for fixed $\Delta=-0.5$ and varying the defect strength from $0$ to $-0.9$ at steps of $0.1$. 
From top to bottom, the absolute strength of the defect $|\delta|$ increases.  
Right: In this panel the defect strength is fixed to $\delta=-0.4$, while we vary the bulk 
anisotropy from $\Delta=-0.6$ (topmost curve) to $\Delta= 1$ (bottommost curve) at steps of  $0.1$. 
The dashed line corresponds to the XX chain $\Delta=0$.}
\end{figure}

The data for the von Neumann entanglement entropy in spin chains of length $L=64$ as function of time are shown 
in Fig. \ref{fig64}, where in the left panel we fix $\Delta=-0.5$ and report all the values of $\delta$, while 
in the right panel we fix $\delta=-0.4$ and show all the considered values of $\Delta\in [-0.6,1]$ up to $t=2L/v$. 

For $L=128$, the simulations are computationally more demanding. Thus we focus on few most significative  
values of $\Delta$ and $\delta$ and we only study the non-equilibrium dynamics up to the revival time
$t_r=L/v$ since the universal part of the evolution should be periodic with period $t_r$.
The simulations allow us to draw the following general scenario which we anticipate before the 
careful analysis of the data.
The non-equilibrium behavior after the local quench is reminiscent of the equilibrium one: 
the entropies evolve in time according to the CFT prediction (\ref{S_CFT}), 
provided that  $\Delta<0$, i.e. attractive interaction;
conversely, for repulsive interactions ($\Delta>0$), the entanglement growth is suppressed and the suppression increases with 
repulsive coupling constant $\Delta$. 
Thus, as for the equilibrium counterpart, there is a fundamental difference between repulsive and attractive interactions: 
even out-of-equilibrium, the operator associated to the impurity turns out to be irrelevant for $\Delta<0$ and relevant for $\Delta>0$. 
Obviously, the `velocity of the RG flow' toward its asymptotic limit strongly depends on the defect strength.
In the following we report and discuss the results for Von Neumann and R\'enyi entropies to support this scenario 
in two separate subsections.

\subsection{Von Neumann entropy}

\begin{figure}[t]
\includegraphics[width=0.5\textwidth]{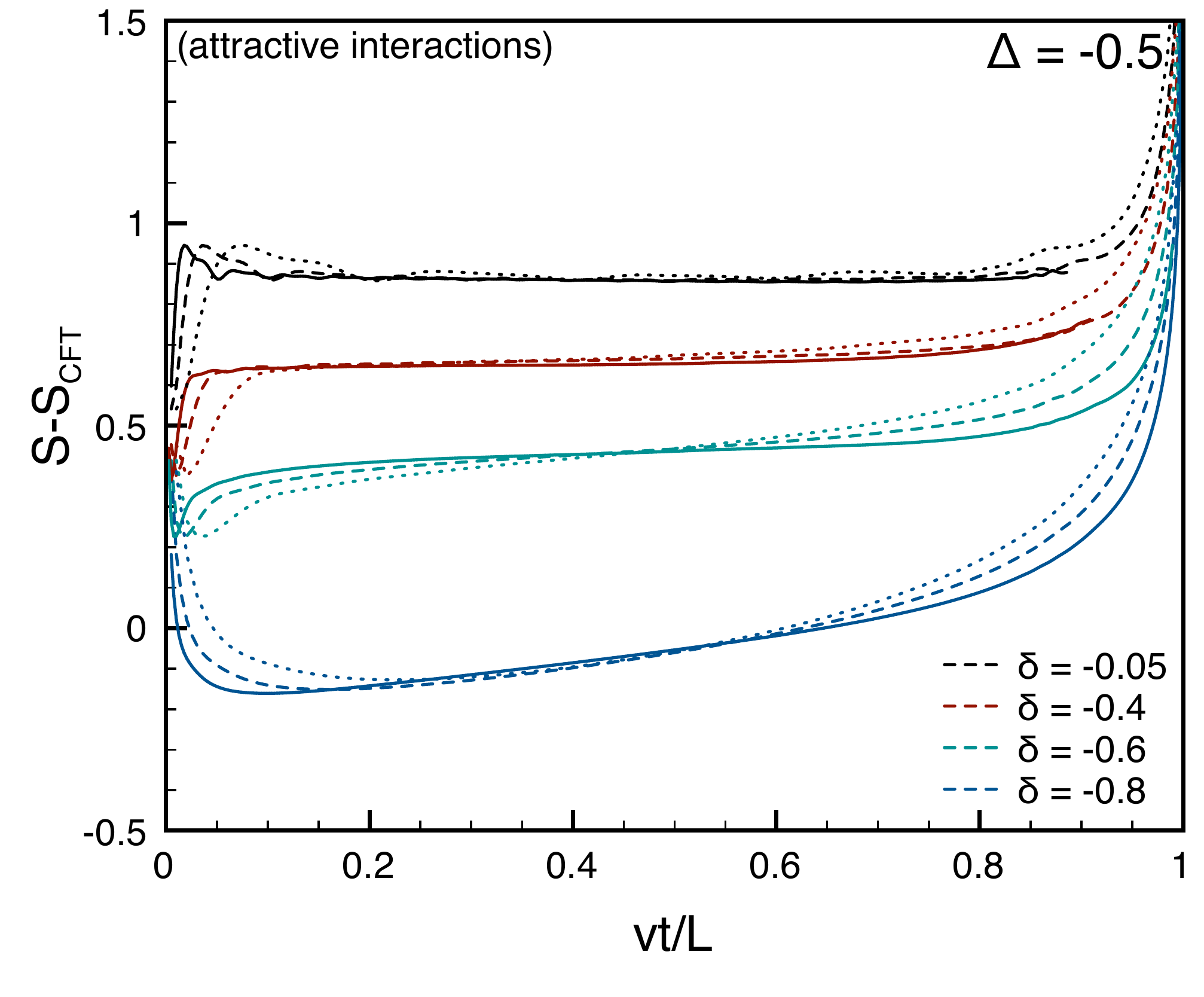}\includegraphics[width=0.5\textwidth]{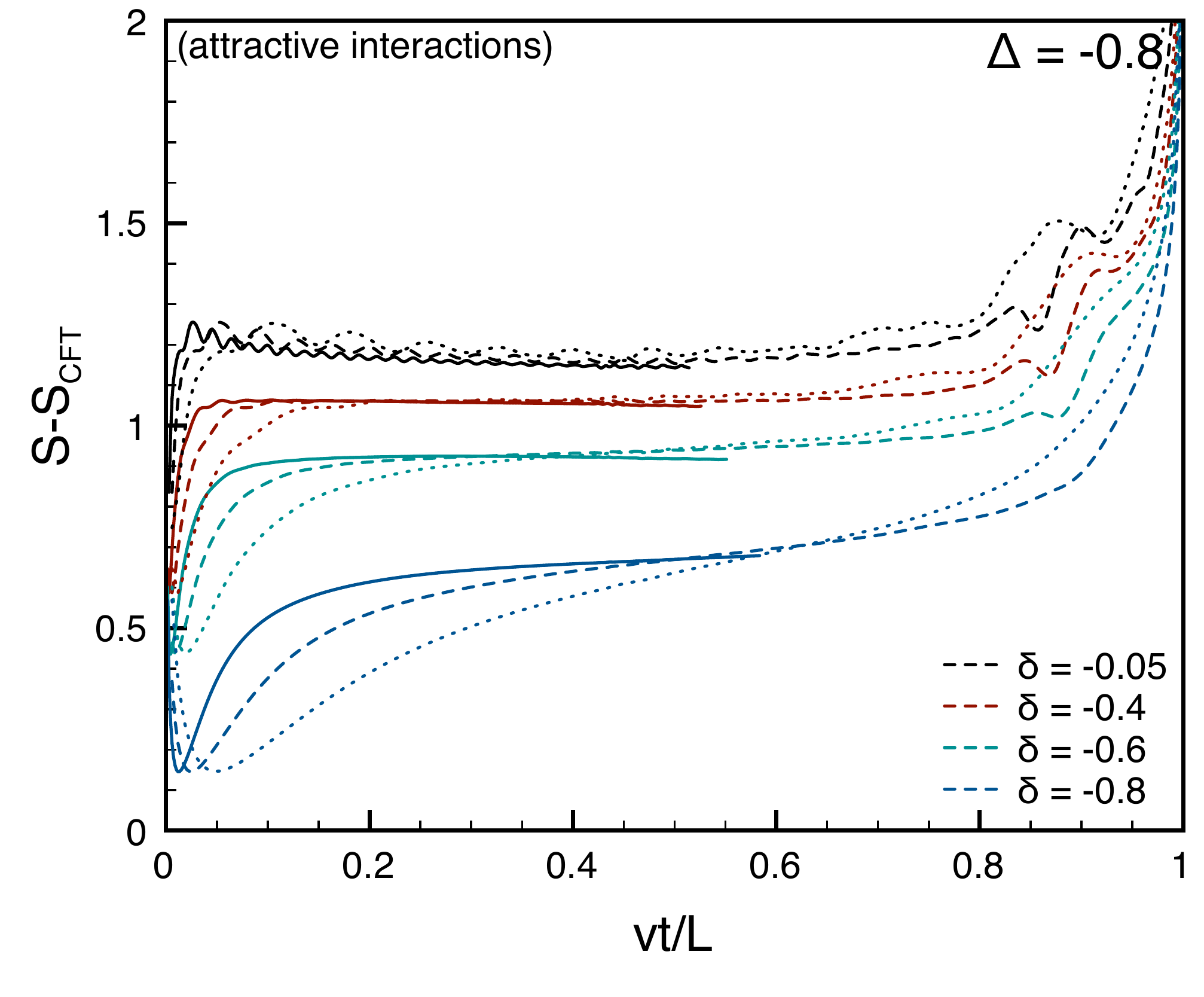}
\caption{\label{fig2} 
Differences between the entanglement entropy evolution and the CFT prediction (\ref{S_CFT}) 
in the attractive regime ($\Delta<0$) 
for two different values of $\Delta$ and four representative values of the defect bond. 
Dotted lines are for $L=32$, dashed lines for $L=64$, and full lines for $L=128$.
Increasing the system size these differences tend to flatten, showing that, for 
$\Delta<0$, a defect $\delta\in(-1,0)$ is always irrelevant.}
\end{figure}

We now report the numerical data for the Von Neumann entanglement entropy discussing separately 
the cases of  attractive and repulsive interactions. 
Fig. \ref{fig2} shows the results for attractive interaction $\Delta<0$ (note that for $\Delta=-0.8$ the data for $L=128$
do not reach the revival time $L/v$ because this would have been computationally too demanding since $v$ is very small
cf. Eq. (\ref{v_F})). 
In order to show accurately that, in the thermodynamic limit, independently of the defect strength $\delta\in(-1,0)$, 
the entanglement entropy is described by the CFT prediction (\ref{S_CFT}),
we have subtracted the expected conformal formula (\ref{S_CFT}) to the time evolution of the entanglement entropy. 
For small ($\delta=-0.05$) and moderate ($\delta=-0.4$)  defects, the difference $S-S_{CFT}$ clearly tends to a constant  
value (independent on $t$ and $L$) as long as the rescaled variable $vt/L$ is sufficiently far from the boundaries 
$vt/L\sim 0$ or $1$, where finite size effects are enhanced. 
However, when the defect gets stronger (e.g. $\delta=-0.8$) deviations from constant behavior are evident. 
Increasing the system sizes, these curves tend to flatten for both values of the anisotropic parameter reported in the figure,
suggesting that in the thermodynamic limit the difference would be perfectly constant for any $\Delta$. 
A proper finite-size scaling analysis is made difficult by the non-uniform approach as function of  $vt/L$, 
but already a qualitative look of the data leaves no doubts on the asymptotic result. 
Notice that for $\Delta=-0.5$ larger system sizes are required for the differences to get constant compared to 
$\Delta=-0.8$. Fig. \ref{fig64} shows that this is a general trend as function of $\Delta$, i.e. smaller is 
$|\Delta|$ larger is the system size required to observe the conformal behavior. 
The origin of this dependence on $\Delta$ is easily understood. 
For $\Delta=0$, the time evolution 
of the entanglement entropy is not given by the CFT formula  (\ref{S_CFT}), but the logarithm 
has a defect dependent prefactor \cite{eisler2012}.
Thus, for small $\Delta$, increasing $L$ there is a crossover between the $\Delta=0$ behavior 
and the asymptotic result.

In the case of repulsive interactions (XXZ antiferromagnet with $\Delta>0$), let us first discuss what we 
would expect if, in analogy with equilibrium, the defect would be relevant. 
For any $\Delta$, the entanglement entropy should initially grow, but at some given time $t^*$
the presence of the relevant defect prevents further passage of information between the two halves and 
the entanglement saturates to a constant value (up to a possible revival at $L/v$). 
Clearly  $t^*$ is not universal and its value would depend  both on $\delta$ and $L$. 
When $\delta$ is small enough, we expect this saturation time to be so large 
that for any accessible size $L$, it would be very difficult to find a perceptible difference with 
the CFT behavior at $\delta=0$. 
Increasing $\delta$, the saturation time $t^*$ decreases, even if for small and moderate values of $L$,
the difference with the CFT evolution should remain small. 
Further increasing $\delta$, the saturation time $t^*$ should become very short and 
the entanglement quickly saturates to its asymptotic value.

\begin{figure}[t]
\includegraphics[width=0.5\textwidth]{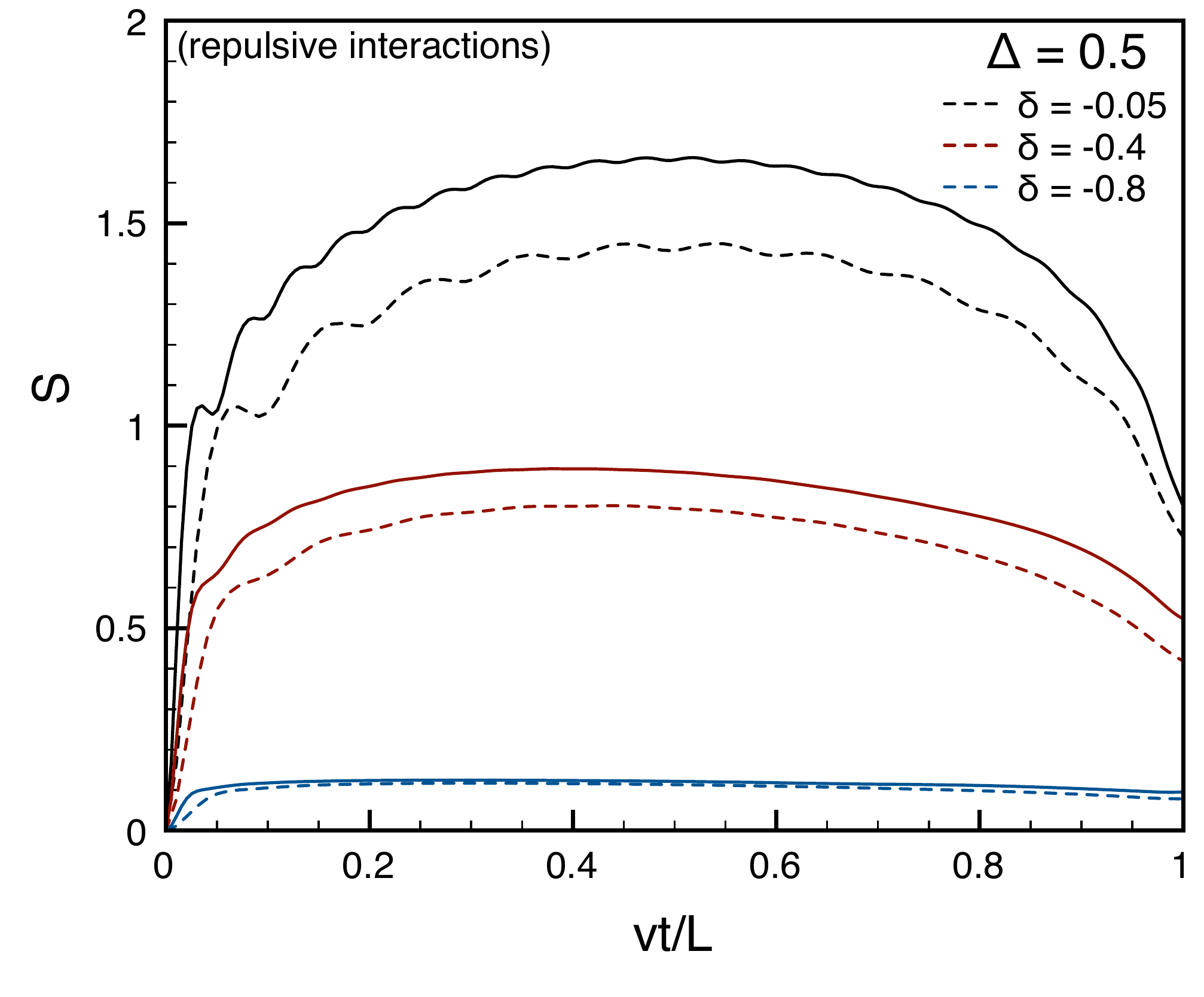}\includegraphics[width=0.5\textwidth]{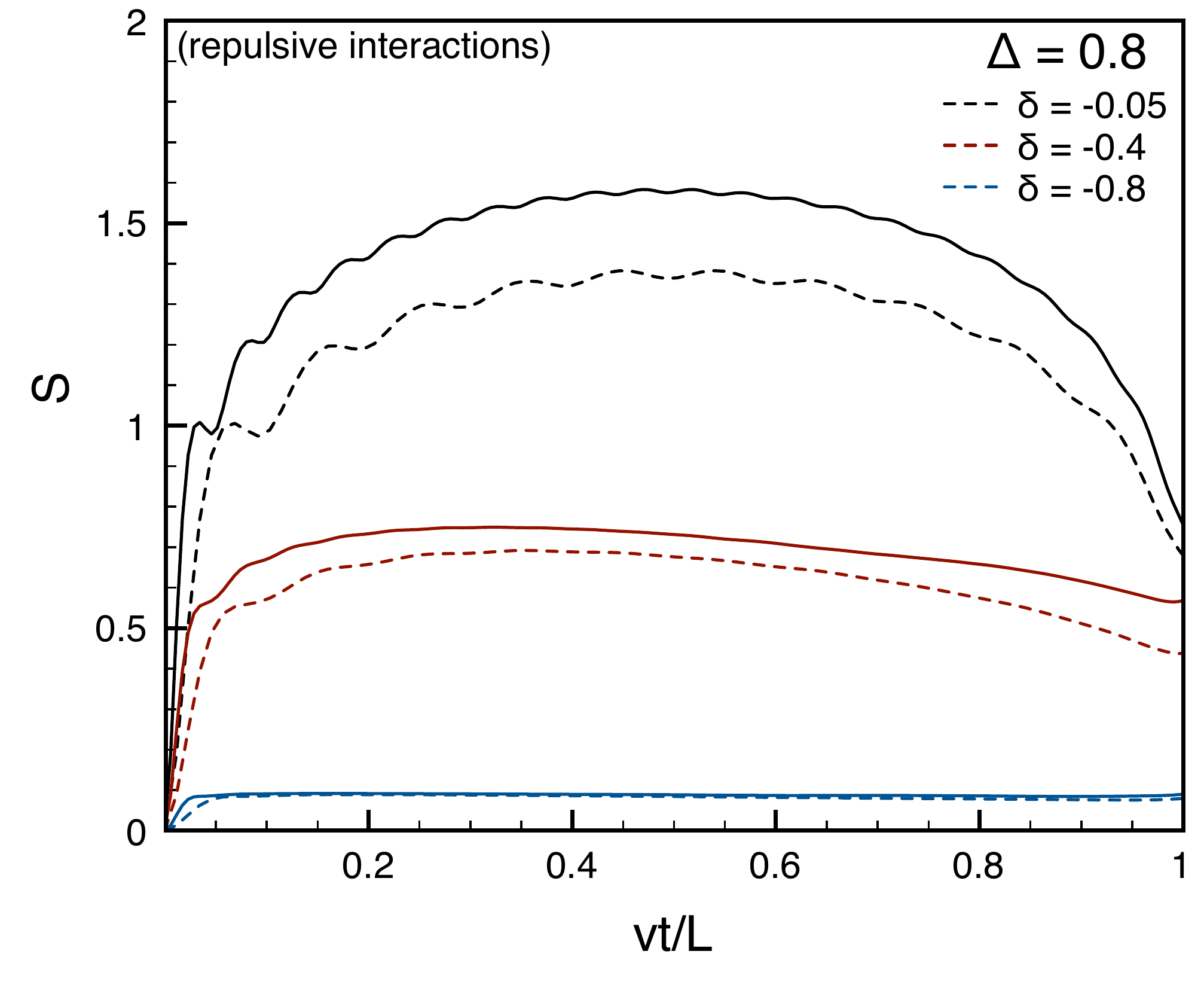}
\caption{\label{fig3} 
Entanglement entropy evolution in the repulsive regime: 
the entanglement initially grows and then saturates to a time independent value, 
but the saturation time is very large for small $\delta$ and would 
require very large systems to be observed.
Dashed lines are for $L=64$ and full lines for $L=128$.}
\end{figure}
 
If this scenario is correct, it does not make  sense to subtract the conformal formula to tDMRG data, 
and so we simply report the entanglement entropy evolution for two values of $\Delta$ in  Fig.~\ref{fig3}, 
supplementing the data in the left panel of Fig. \ref{fig64} for $L=64$.
From these two figures, it is clear that the previously conjectured scenario for the time evolution is confirmed by the numerics.  
For all values of $\Delta$, the growth of the entanglement entropy  
observed in homogeneous systems is more and more suppressed as the absolute value of the  
defect strength increases.  
For the smallest impurity we considered ($\delta=-0.05$), the entanglement entropy is very close to the conformal 
prediction, but the lower quality of the fit (that we do not report in the plot) compared to the data at $\delta=0$ 
suggests that some tiny differences can be already present. 
As the defect becomes stronger (e.g. $\delta=-0.4$), the curves flatten substantially and slowly approach an 
horizontal line in the thermodynamic limit. 
For very strong defect ($\delta=-0.8$) the entanglement entropy becomes flat in a very short time. 
Of course, also in this case the bulk interaction plays a role because  
the saturation limit is approached faster for larger values of $\Delta$.
As in the case of attractive interactions, this is due to the fact that for $\Delta=0$ there is a logarithmic growth of the 
entanglement for any $\delta$ and so there is a crossover in $L$ from the non-interacting 
behavior to the saturation in the interacting chain. 

Finally we also discuss the revival of the entanglement entropy. 
In the right panel of Fig. \ref{fig64}, we show the time evolution up to $2 L/v$ (i.e. for two full asymptotic periods) 
for $L=64$, $\delta=-0.4$ and several values of $\Delta$.
For attractive interaction $\Delta<0$, the quasi-periodic behavior is evident, 
but during the second period some oscillations in time are superimposed to the CFT prediction (\ref{S_CFT}).
These have been already observed for $\Delta=0$ (see e.g. \cite{ds-11}), even in the absence of the defect,
and they are mainly due the interference between excitations with non perfectly equal velocities.
It has been shown \cite{ds-11} that the amplitude of these oscillations goes to zero increasing the system size 
for $\Delta=0$ and $\delta=0$. 
From the figure it is evident that for $L=64$, the amplitude of the oscillations gets smaller decreasing $\Delta$,
thus we expect they should vanish in the thermodynamic limit for all $\Delta<0$, as 
also corroborated by the data for $L=32$ not shown here. 
For $\Delta>0$, the effect of the slower quasi-particles is different.
Indeed, Fig. \ref{fig64} shows that  the interference effect gives rise to 
oscillations which make the curve flatter in time during the second period.
Thus the second period entanglement entropy is 
closer to the thermodynamic limit compared to the first period value. 
Notice in fact that in Fig. \ref{fig64} for $\Delta>0$ and fixed $\delta=-0.4$, 
the average over the second period is slightly larger than the average 
over the first one, similarly as in Fig. \ref{fig3} the entanglement for $L=128$ 
is larger that the one for $L=64$ at $\delta=-0.4$.

\begin{figure}[t]
\includegraphics[width=\textwidth]{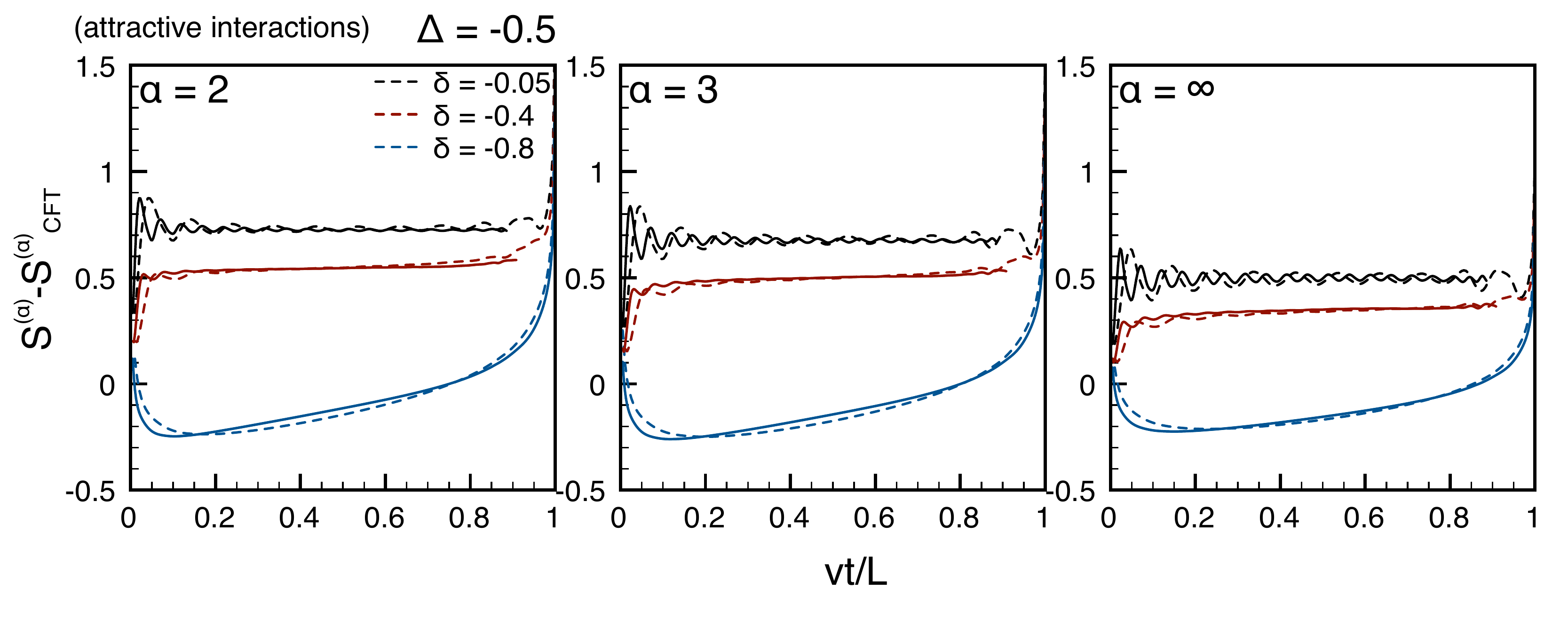}\\
\includegraphics[width=\textwidth]{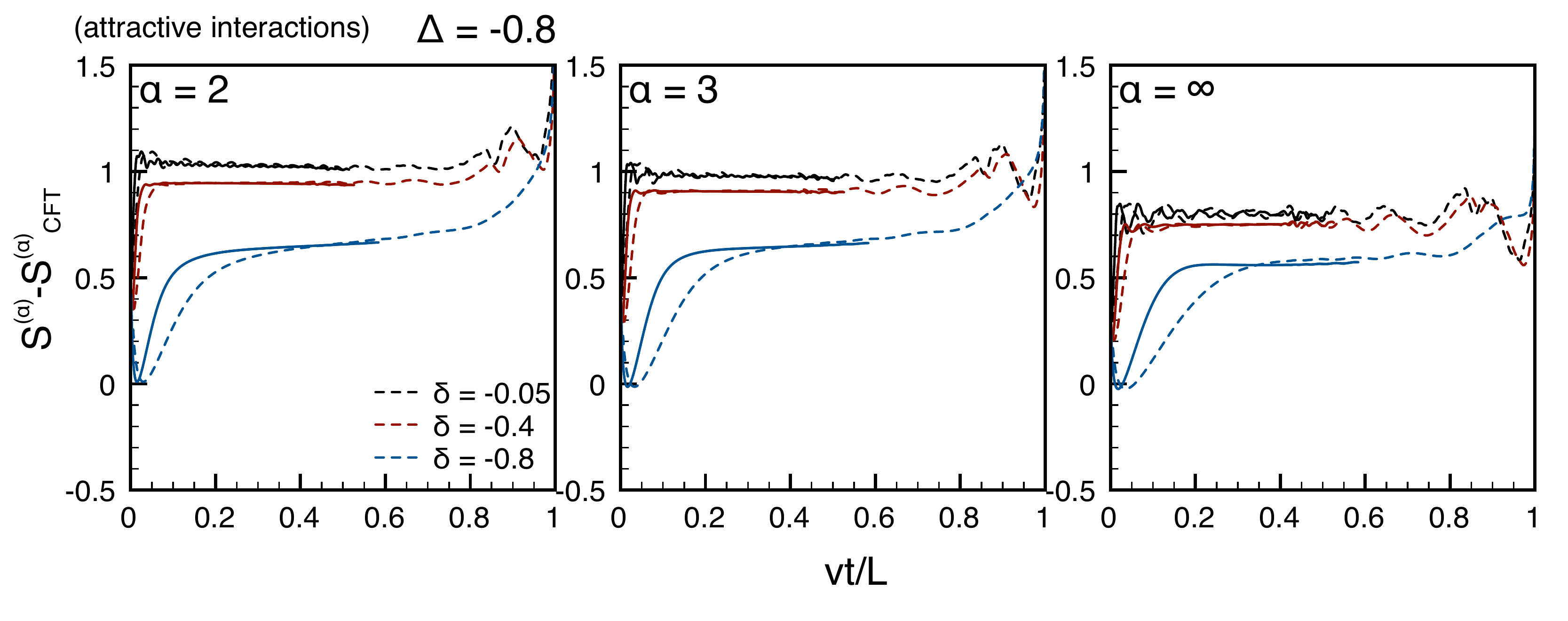}
\caption{\label{fig4} 
Differences between the R\'enyi entropy evolution and the CFT prediction in the attractive regime for three
different values of $\alpha$. Dashed lines are for $L=64$ and full lines for $L=128$.
We do not report the data for $L=32$ to make the plots more readable.}
\end{figure}

\subsection{R\'{e}nyi entropies}

In this section we check that the general scenario drawn above on the basis of the von Neumann entropy  carries over to 
the time evolution of the R\'enyi entropies for general $\alpha$, limiting to report data for $\alpha=2,3,\infty$ (the case
$\alpha=\infty$ corresponds to the logarithm of the largest eigenvalues of the reduced density matrix, i.e. the so-called 
single copy entanglement \cite{sce}).

\begin{figure}[t]
\includegraphics[width=\textwidth]{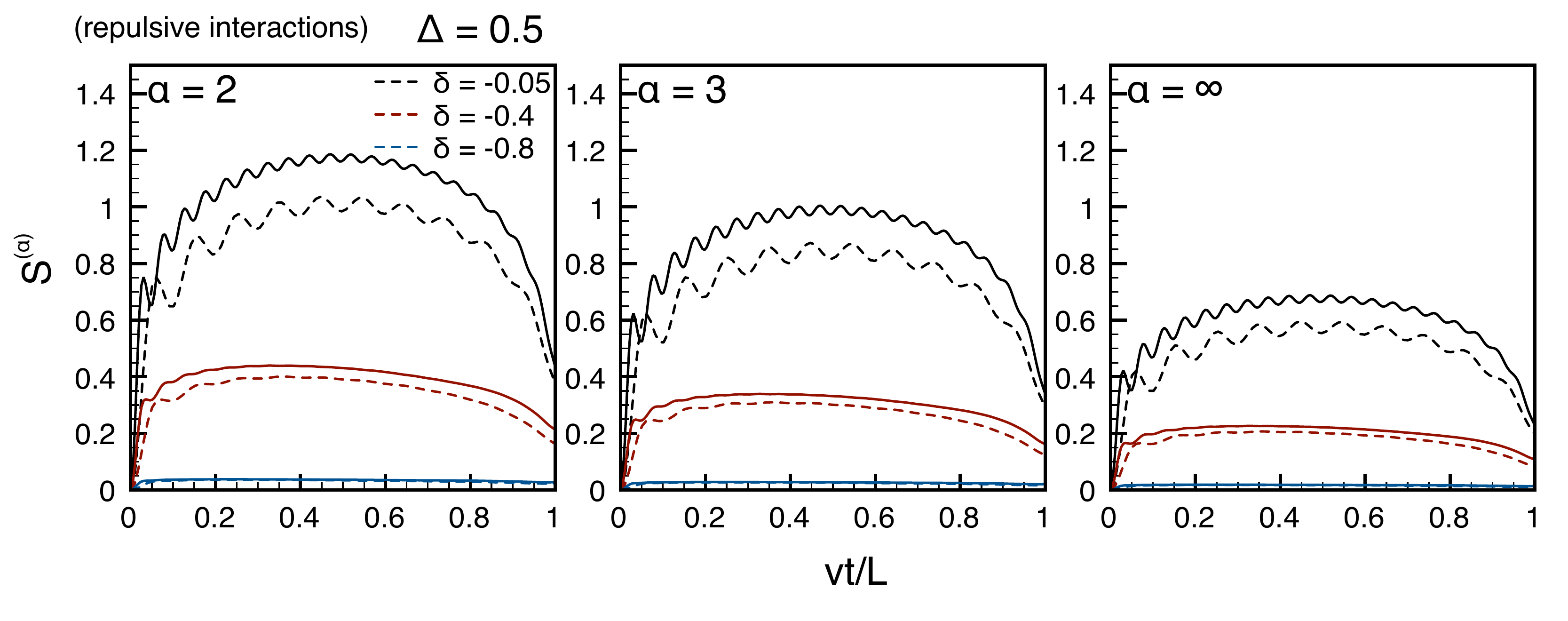}\\
\includegraphics[width=\textwidth]{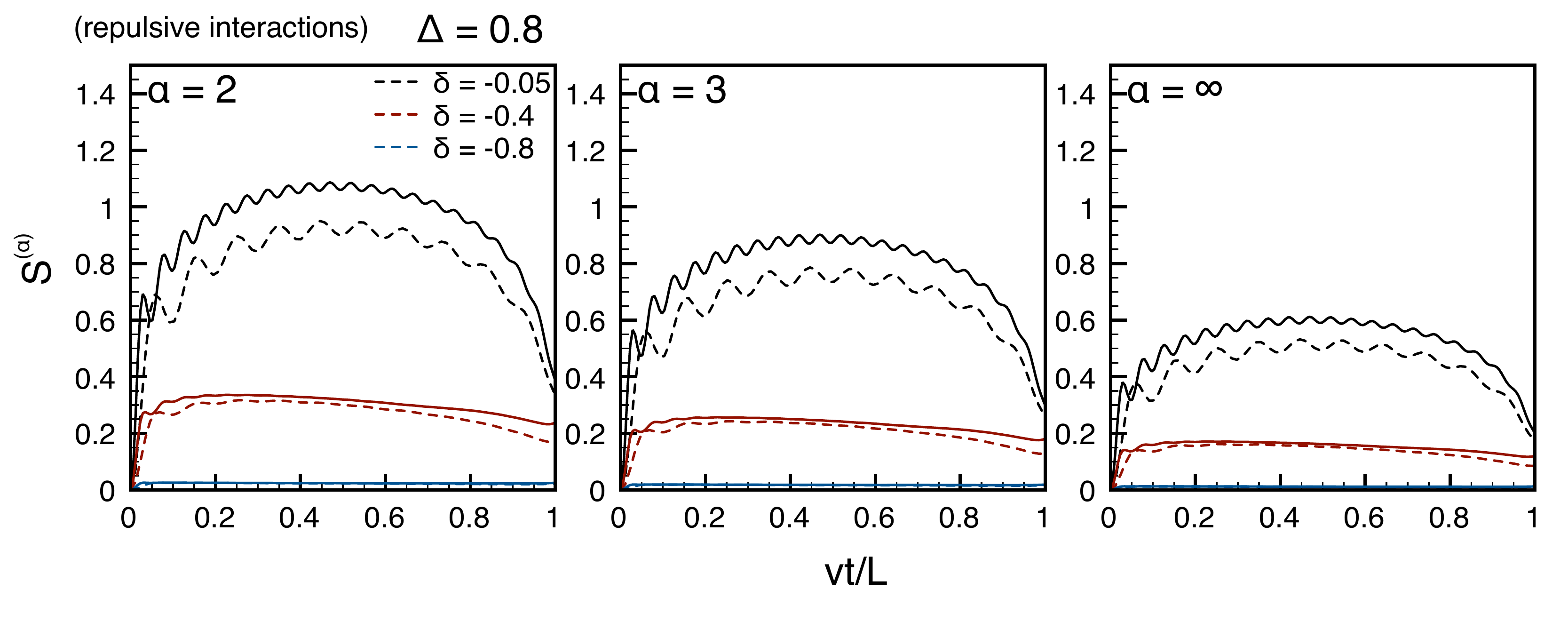}
\caption{\label{fig5} The evolution of the R\'enyi entropies in the repulsive regime for different value of $\alpha$. 
Dashed lines are for $L=64$ and full lines for $L=128$.}
\end{figure}

In Fig. \ref{fig4} we report the results for the attractive case $\Delta<0$. 
In analogy with the von Neumann entropy, we plot the difference $S^{(\alpha)}- S^{(\alpha)}_{CFT}$.
For all values of $\alpha$, $\Delta$, and $\delta$, the irrelevance of the defect is confirmed by these data,
indeed for small $|\delta|$ the CFT prediction is already approached even for relatively small values of $L$, 
while for large $|\delta|$ the difference tends to a constant only in the limit of large $L$. 
However, there is also an unexpected effect. Indeed, in Fig. \ref{fig4}, for small value of 
$\delta$, there are subleading oscillations on top of the leading CFT behavior. 
The presence of these oscillations is well-known and it has been observed in the absence of the 
defect in all previous studies (see e.g. \cite{ep-08,edpp-08,ds-11}) and it is explained as the non-equilibrium 
counterpart of the equilibrium unusual corrections to the scaling \cite{unusual,lsca-06} of the form $L^{-K/n}$, 
with $K$ the Luttinger parameter.
The frequency of these oscillations is related to $k_F$ (i.e. the filling) and does not dependent on the 
coupling strength $K$ \cite{unusual}. 
The interesting new effect is that in the presence of the defect such corrections are largely suppressed 
and almost disappear already for $\delta=-0.4$. 
This means that the CFT predictions becomes more accurate in the presence of a defect (that however 
must be not too large to observe the flattening of the curves in Fig. \ref{fig4}).
We stress that these oscillations are not directly related to the ones discussed in the previous section due to 
quasi-particles with velocity different from $v$, which in Fig. \ref{fig4} starts manifesting only close to the 
revival time.

The data for the repulsive case are reported in Fig. \ref{fig5}. 
Even in this case, the picture drawn from the study of the von Neumann entropy straightforwardly 
carries over to the R\'enyi: 
for the small impurities, the R\'enyi entropies stay very close to the conformal 
prediction, but as the defect becomes stronger, the entropies flatten and approach quickly a saturation value 
for any $\alpha$. 
Even in the repulsive case, it is evident that the oscillating corrections to the scaling present for small $|\delta|$
are largely suppressed as the defect strength increases.

\section{Conclusions}

In this manuscript, we have studied the time evolution of the entanglement entropies 
after a local quench connecting two anisotropic spin-1/2 XXZ Heisenberg chains 
via an impurity bond. 
By means of tDMRG, we showed that, at half-filling, the asymptotic behavior of the entanglement entropies for 
large systems depends only on the sign of bulk interactions. 
For attractive interaction ($\Delta<0$), it turns out that the defect is irrelevant and the 
evolution is asymptotically equivalent to the conformal one without the impurity.
For repulsive bulk interaction ($\Delta>0$), the defect becomes relevant and the 
entanglement entropy saturates to a finite value after a finite time 
which are both not universal and depend on the defect strength.

We have only determined time evolution of the entanglement entropy between the two initially disconnected parts, 
but we can safely conclude
that the same picture should carry over to all other observables which can be determined just 
by exploiting the relevance/irrelevance of the defect according to the sign of the bulk interaction.
We mention that the correspondence between the equilibrium and out of equilibrium situations 
resulting from a local perturbation for observables different from the entanglement entropies
has a long history \cite{al-94,az-97} such as the connection between equilibrium Anderson orthogonality catastrophe 
and the X-ray absorption spectra (corresponding to a local perturbation). 
Furthermore, while we have only considered the XXZ spin chain, in view of the Luttinger liquid universality, 
the same conclusions apply to all other models in the same universality class, such as
one-dimensional gases of spinless bosons (both the continuum Lieb-Liniger and the Bose-Hubbard model in the
superfluid phase), quantum wires junction etc. 
Oppositely, for multi-component continuum or lattice models of interacting fermions (such as the Hubbard model)
or Fermi-Bose mixtures, the phenomenon of spin-charge separation could lead to different universal 
features, e.g. reminiscent of what observed in Ref. \cite{kkmgs-08} 
for different initial states.
For these multi-component models, also 
shell-filling effects \cite{elc-13} could strongly affect the non-equilibrium dynamics 
of the entanglement entropy. 
Therefore it would be really interesting to investigate local quenches in these 
models both in the absence and in the presence of defect bonds.

\section*{Acknowledgments}
Both authors acknowledge the financial  support by the ERC under  Starting Grant  279391 EDEQS.

\end{document}